\documentclass[12pt,a4,onecolumn,secnumarabic,amssymb, amsmath, nofootinbib,tightenlines,
nobibnotes, aps, prl,epsfig]{revtex4}
\usepackage{graphicx}
\usepackage{dcolumn}
\usepackage{bm}
\begin{document}
\preprint{APS/123-QED}
\title{ Solutions of independent DGLAP  evolution equations for the gluon distribution and singlet structure
functions in the next-to-leading order analysis at low $x$ }

\author{G.R.Boroun}
\altaffiliation{boroun@razi.ac.ir}
\author{}%
\affiliation{ Physics Department, Razi University, Kermanshah
67149, Iran}
\date{\today}
\baselineskip=30pt
\begin{abstract}
We  present a set of independent formulae to extract the gluon
distribution  and  singlet structure function from its derivatives
with respect to $lnQ^{2}$ in the next- to- leading order of
perturbation theory at low-$x$ based on a hard pomeron exchange.
In this approach, both singlet quarks and gluons have the same
high-energy behavior at small $x$. This approach needs the QCD
input parameterizations for a independent DGLAP evolutions that we
calculated numerically and compared with MRST, GRV and DL model-
the Pomeron has a hard nature. Its evolution gives a good fit to
the experimental data. The obtained values are in the range
$10^{-4}\hspace{0.1cm}{\leq}\hspace{0.1cm}x\hspace{0.1cm}{\leq}10^{-2}$
at $Q^{2}=20 \hspace{0.1cm}GeV^{2}$.
\end{abstract}
\maketitle
\newpage
\subsection{1 Introduction}

The DGLAP $[1]$ evolution equations are fundamental tools to study
the $Q^{2}$ and $x$ evolutions of structure functions, where $x$
and $Q^{2}$ are Bjorken scaling and four momenta transfer in deep
inelastic scattering (DIS) process respectively $[2]$. The
measurements of the $F_{2}(x,Q^{2})$ structure functions by DIS
processes in the small- $x$ region, have opened a new era in
parton density measurements inside hadrons. The structure function
reflects the momentum distributions of the partons in the nucleon.
It is also important to know of the gluon distribution inside a
hadron at low- $x$ since gluons are expected to be dominant in
this region. The steep rise of $F_{2}(x,Q^{2})$ towards low $x$
observed at HERA, also indicates in perturbative quantum
chromodynamics (PQCD) a similar rise of the gluon distribution
towards low $x$. In the usual procedure the DIS data are analyzed
by the NLO QCD fits based on the numerical solution of the DGLAP
evolution equations and it is found that the DGLAP analysis can
well describe the data in the perturbative region $Q^{2}{\geq}1
GeV^{2}$ [3]. Alternative to the numerical solution, one can study
the behavior of the quarks and gluons through the analytical
solutions of the evolution equations. Although exact analytical
solutions of the DGLAP equations are not possible in the entire
range of $x$ and $Q^{2}$, but under certain conditions analytical
solutions are possible [4-5] which are quite successful as far as
the HERA small $x$ data are concerned.\\

Small $x$ behavior of structure functions for fixed $Q^{2}$
reflects the high energy behavior of the virtual Compton
scattering total cross section with increasing total CM energy
squared $W^{2}$ since $W^{2}=Q^{2}(1/x-1)$. The appropriate
framework for the theoretical description of this behavior is the
Regge pole exchange picture [6]. It can be asserted confidently
that Regge theory is one of the most successful approaches to
describe high energy scattering of hadrons. This high energy
behavior can be described by two contributions: an effective
Pomeron with its intercept slightly above unity ($\sim$1.08) and
the leading meson Regge trajectories with intercept
$\alpha_{R}(0){\approx}0.5$ [7].\\

The Regge pole model gives the following parametrization of the
deep inelastic scattering structure function $F_{2}(x,Q^{2})$ at
small $x$:
\begin{equation}
F_{2}(x,Q^{2})=\sum_{i}\widetilde{\beta}_{i}(Q^{2})x^{1-\alpha_{i}(0)},
\end{equation}

Where the singlet part of the structure function $F_{2}$ is
controlled at small $x$ by Pomeron exchange, while the non-singlet
part
$F_{2}^{NS}=F_{2}^{p}-F_{2}^{n}$ by the $A_{2}$ reggeon [3].\\

At small $x$ the dominant role is played by the gluons and the
basic dynamical quantity is the unintegrated gluon distribution
$f(x,Q_{t}^{2})$ where $x$ denotes the momentum fraction of a
parent hadron carried by a gluon and  $Q_{t}$ its transverse
momentum. The unintegrated distribution $f(x,Q_{t}^{2})$ is
related in the following way to the more familiar scale dependent
gluon distribution $xg(x,Q^{2})$ [4]:
\begin{equation}
xg(x,Q^{2})=\int^{Q^{2}}\frac{dQ_{t}^{2}}{Q_{t}^{2}}f(x,Q_{t}^{2}).
\end{equation}
In the leading $ln(1/x)$ approximation the unintegrated
distribution $f(x,Q_{t}^{2})$ satisfies the BFKL equation [8]
which has the following form:
\begin{eqnarray}
f(x,Q_{t}^{2})&=&f^{0}(x,Q_{t}^{2})+\overline{\alpha}_{s}\int_{x}^{1}\frac{dx'}{x'}\int\frac{d^{2q}}{{\pi}q^{2}}[\frac{Q_{t}^{2}}{(\mathbf{q}+\mathbf{Q_{t}})^{2}}\nonumber\\
&&f(x',(\mathbf{q}+\mathbf{Q_{t}})^{2})-f(x',Q_{t}^{2})\Theta(Q_{t}^{2}-q^{2})],\nonumber\\
\end{eqnarray}
where
\begin{equation}
\overline{\alpha}_{s}=\frac{3\alpha_{s}}{\pi}.
\end{equation}

This equation sums over the ladder diagrams with gluon exchange
accompanied by virtual corrections which are responsible for the
gluon reggeization. For the fixed coupling case, this equation can
be solved analytically and the leading behavior of its solution at
small $x$ is given by the following expression:
\begin{eqnarray}
f(x,Q_{t}^{2}){\sim}(Q_{t}^{2})^{\frac{1}{2}}\frac{x^{-\delta_{BFKL}}}
{\sqrt{ln(\frac{1}{x})}}\hspace{0.1cm}exp(-\frac{ln^{2}(Q_{t}^{2}/\overline{Q}^{2})}{2\lambda^{,,}ln(1/x)})
\end{eqnarray}
with $\lambda_{BFKL}=4ln(2)\overline{\alpha}_{s}$ and
$\lambda^{,,}=\overline{\alpha}_{s}28\zeta(3)$. Where the Riemann
zeta function $\zeta(3){\approx}1.202$. The parameter
$\overline{Q}$ is of nonperturbative origin.\\

The quantity $1+\lambda_{BFKL}$ is equal to the intercept of the
so-called BFKL Pomeron. Its potentially large magnitude
(${\sim}$1.5) should be contrasted with the intercept
$\alpha_{soft}{\approx}1.08$ of the effective soft Pomeron which
has been determined from the phenomenological analysis of the high
energy behavior of hadronic and photoproduction total cross
sections. When the model [7] was applied in deep inelastic
scattering, namely to the proton structure functions, one needs to
add a second Pomeron, "hard" (in contrast with the first one
called a "soft" Pomeron, because of its intercept near 1),
with a larger intercept $\alpha_{hp}{\approx}1.4$ [9,10].\\

The hypothesis of the Pomeron with data of the total cross section
shows that a better description is achieved in alternative models
with the Pomeron having intercept one, but with a harder $j$
singularity (a double pole) [11]. This model has two Pomeron
components, each of them with intercept $\alpha_{P}=1$; one is a
double pole and the other one is a simple pole [12].\\

 One is, however, tempted to explore the possibility of
obtaining approximate analytical solutions of DGLAP equations
themselves at least in the restricted domain of low- $x$.
Approximate solutions of DGLAP equations have been reported
$[13-15]$ with considerable phenomenological success. In such an
approximate scheme, one uses a Taylor expansion valid at low- $x$
and reframes the DGLAP equations as partial differential equations
in the variable $x$ and $Q^{2}$ which can be solved by standard
methods.\\

In this paper we  suggest an approximate analytical independent
solutions of the next- to- leading order (NLO) DGLAP equations for
the gluon distribution and the singlet structure function,
respectively.  Therefore we concentrate on the Pomeron in our
calculations, although clearly good fits relative to results show
that the gluon distribution and the singlet structure function
need a model having hard Pomeron. We compare our results with the
exacted ones GRV98[16], MRST2001[17] and  DL fit[10] parton
distributions. Our paper is organized as follows. In section $2$
solutions of the DGLAP equations by the Taylor
expansion are presented while section $3$ is devoted to results  and discussions.\\
\subsection{2 Solution of the DGLAP equations by the Taylor Expansion}

The HERA data should determine the small $x$ behavior of gluon and
singlet quark distributions. We will be concerned specifically
with the singlet contribution to the proton structure function:
\begin{eqnarray}
F^{ep}_{2}(x,Q^{2})=\frac{5}{18}\Sigma(x,Q^{2})+\frac{3}{18}F^{NS}_{2}(x,Q^{2})\\\nonumber
\Sigma(x,Q^{2}){\equiv}x\sum_{i=1}^{N_{f}}(q_{i}(x,Q^{2})+\overline{q}_{i}(x,Q^{2})),
\end{eqnarray}
where $N_{f}$ is the number of active flavors. At small $x$ the
nonsinglet contribution $F^{NS}_{2}(x,Q^{2})$ is negligible and
can be ignored. At small $x$ and large $Q^{2}$ the singlet quark
distribution $\Sigma(x,Q^{2})$ is essentially driven by the
generic instability of the gluon distribution  $xg(x,Q^{2})$. To
see how this works, consider the singlet Altarelli- Parisi
equations [1], which describe perturbative evolution of
$xg(x,Q^{2})$ and
$\Sigma(x,Q^{2})$.\\

The DGLAP evolution equations for the singlet quark structure
function and the gluon distribution have the forms:
\begin{widetext}
\begin{equation}
\frac{dG(x,Q^{2})}{dlnQ^{2}}=\frac{\alpha_{s}}{2\pi}{\int_{0}^{1-x}}dz[
P^{LO+NLO}_{gg}(1-z) G(\frac{x}{1-z},Q^{2})+P^{LO+NLO}_{gq}(1-z)
\Sigma(\frac{x}{1-z},Q^{2})]
\end{equation}
\begin{equation}
\frac{d\Sigma(x,Q^{2})}{dlnQ^{2}}=\frac{\alpha_{s}}{2\pi}{\int_{0}^{1-x}}dz[
P^{LO+NLO}_{qq}(1-z)
\Sigma(\frac{x}{1-z},Q^{2})+2n_{f}P^{LO+NLO}_{qg}(1-z)
G(\frac{x}{1-z},Q^{2})]
\end{equation}
\end{widetext}
where the splitting functions are the LO and NLO Altarelli- Parisi
splitting kernels [1,18]. The running coupling constant
$\frac{\alpha_{s}}{2\pi}$ has the form in the NLO as:
\begin{equation}
\frac{\alpha_{s}}{2\pi}=\frac{2}{\beta_{0}t}[1-\frac{\beta_{1}lnt}{\beta_{0}^{2}t}]
\end{equation}
with $\beta_{0}=\frac{1}{3}(33-2N_{f})$ and
$\beta_{1}=102-\frac{38}{3}N_{f}$. The variable $t$ is defined as
$t=ln(\frac{Q^{2}}{\Lambda^{2}})$ and the $\Lambda$ is the QCD
cut- off parameter.\\

To find an analytic solution, we note that the splitting kernels
as $z{\rightarrow}\hspace{0.1cm}0$ have the following forms [19]:
\begin{eqnarray}
P^{LO+NLO}_{gg}(z)&=&\frac{2C_{A}}{z}+\frac{\alpha_{s}}{2\pi}\frac{(12C_{F}N_{f}T_{R}-46C_{A}N_{f}T_{R})}{9z},\nonumber\\
P^{LO+NLO}_{gq}(z)&=&\frac{2C_{F}}{z}+\frac{\alpha_{s}}{2\pi}\frac{(9C_{F}C_{A}-40C_{F}N_{f}T_{R})}{z},\hspace{0.6cm}\nonumber\\
P^{LO+NLO}_{qq}(z)&=&\frac{\alpha_{s}}{2\pi}\frac{40C_{F}N_{f}T_{R}}{9z},\hspace{3.6cm}\nonumber\\
P^{LO+NLO}_{qg}(z)&=&\frac{\alpha_{s}}{2\pi}\frac{40C_{A}N_{f}T_{R}}{9z}.\hspace{3.6cm}
\end{eqnarray}
For an SU(N) gauge group we have $C_{A}=N$, $C_{F}=(N^{2}-1)/2N$,
 $T_{F}=N_{f}T_{R}$, and $T_{R}=1/2$, that $C_{F}$ and $C_{A}$ are the color Cassimir operators.\\

We introduce the standard parameterizations  of gluon and singlet
distribution functions as:
\begin{eqnarray}
\Sigma(x,Q^{2})=A_{S}x^{-\delta_{S}}(1-x)^{\nu_{S}}(1+\epsilon_{S}\sqrt{x}+\gamma_{S}x){\equiv}\widetilde{\Sigma}(x,Q^{2})x^{-\delta_{S}},\nonumber\\
G(x,Q^{2})=A_{g}x^{-\delta_{g}}(1-x)^{\nu_{g}}(1+\epsilon_{g}\sqrt{x}+\gamma_{g}x){\equiv}\widetilde{G}(x,Q^{2})x^{-\delta_{g}}.
\end{eqnarray}
where, the usual assumption is that $\delta_{i(=S,g)}=0$. However,
the small $x$ behavior could well be more singular. Note that the
behavior of Eq.(11) with a $Q^{2}$ independent value for
$\delta_{i(=S,g) }$ obeys the DGLAP equations when
$x^{-\delta_{i(=S,g) }}>>1$[4]. According to Regge theory, the
high energy (low $x$) behavior of both gluons and sea quarks is
controlled by the same singularity factor in the complex angular
momentum plane [6], and so we would expect
$\delta_{S}=\delta_{g}=\delta$, where $\delta$ is taken as a
constant factor throughout the calculation. For the structure
functions we take $\widetilde{f}(x,Q^{2})=x^{\delta}f(x,Q^{2})$ to
be finite at $x=0$ with $\delta$ satisfying
$0{\leq}\delta{\leq}\frac{1}{2}$ [20], i.e.
$\widetilde{G}(x)=x^{\delta}G(x)$ and
$\widetilde{\Sigma}(x)=x^{\delta}\Sigma(x)$. Expanding
$\widetilde{G}(x/1-z)$ and $\widetilde{\Sigma}(x/1-z)$ about
$x=0$, we get:
\begin{eqnarray}
\widetilde{G}(\frac{x}{1-z})=\widetilde{G}(0)+\frac{x}{1-z}\widetilde{G}'(0),\nonumber\\
\widetilde{\Sigma}(\frac{x}{1-z})=\widetilde{\Sigma}(0)+\frac{x}{1-z}\widetilde{\Sigma}'(0).
\end{eqnarray}
In these equations, the assumption is the validity of convergence
and neglecting the higher order terms $O(x^2)$.\\

Inserting  Eqs.(10) and (11) in Eqs.(7) and (8) we will have the
DGLAP equations for the gluon and singlet evolutions at low- $x$ :
\begin{eqnarray}
\frac{dG}{dlnQ^{2}}&=&\frac{\alpha_{s}}{2\pi}{\int_{0}^{1-x}}dz(\frac{\beta}{1-z}+\frac{\alpha_{s}}{2\pi}\frac{\gamma}{9(1-z)})(\frac{x}{1-z})^{-\delta}\times(\widetilde{G}(0)+\frac{x}{1-z}\widetilde{G}'(0)]\nonumber\\
&&+\frac{\alpha_{s}}{2\pi}{\int_{0}^{1-x}}dz(\frac{\eta}{1-z}+\frac{\alpha_{s}}{2\pi}\frac{\theta}{9(1-z)})(\frac{x}{1-z})^{-\delta}\times(\widetilde{\Sigma}(0)+\frac{x}{1-z}\widetilde{\Sigma}'(0)]
\end{eqnarray}
and
\begin{eqnarray}
\frac{d\Sigma}{dlnQ^{2}}&=&\frac{\alpha_{s}}{2\pi}{\int_{0}^{1-x}}dz(\frac{\alpha_{s}}{2\pi}\frac{\zeta}{9(1-z)})(\frac{x}{1-z})^{-\delta}\times(\widetilde{\Sigma}(0)+\frac{x}{1-z}\widetilde{\Sigma}'(0)]\nonumber\\
&&+\frac{\alpha_{s}}{2\pi}{\int_{0}^{1-x}}dz(2n_{f})(\frac{\alpha_{s}}{2\pi}\frac{\xi}{9(1-z)})(\frac{x}{1-z})^{-\delta}\times(\widetilde{G}(0)+\frac{x}{1-z}\widetilde{G}'(0)]
\end{eqnarray}
where $\beta=2C_{A}$,
$\gamma=12C_{F}N_{f}T_{R}-46C_{A}N_{f}T_{R}$, $\eta=2C_{F}$,
$\theta=9C_{F}C_{A}-40C_{F}N_{f}T_{R}$, $\zeta=40C_{F}N_{f}T_{R}$
and $\xi=40C_{A}N_{f}T_{R}$.\\

Solving these equations and taking all these considerations into
account, we found:
\begin{eqnarray}
\frac{dG}{dlnQ^{2}}&=&U_{I}[\frac{{\delta}^{{\delta}-1}}{{|{\delta}-1|}^{\delta}}G(x\frac{\delta}{{|{\delta}-1|}})-\frac{1}{\delta}\widetilde{G}(\frac{\delta}{{|{\delta}-1|}})]\nonumber\\
&&+U_{II}[\frac{{\delta}^{{\delta}-1}}{{|{\delta}-1|}^{\delta}}\Sigma(x\frac{\delta}{{|{\delta}-1|}})-\frac{1}{\delta}\widetilde{\Sigma}(\frac{\delta}{{|{\delta}-1|}})],\hspace{1cm}
\end{eqnarray}
and
\begin{eqnarray}
\frac{d\Sigma}{dlnQ^{2}}&=&V_{I}[\frac{{\delta}^{{\delta}-1}}{{|{\delta}-1|}^{\delta}}\Sigma(x\frac{\delta}{{|{\delta}-1|}})-\frac{1}{\delta}\widetilde{\Sigma}(\frac{\delta}{{|{\delta}-1|}})]\nonumber\\
&&+V_{II}[\frac{{\delta}^{{\delta}-1}}{{|{\delta}-1|}^{\delta}}G(x\frac{\delta}{{|{\delta}-1|}})-\frac{1}{\delta}\widetilde{G}(\frac{\delta}{{|{\delta}-1|}})],\hspace{1cm}
\end{eqnarray}
where
$U_{I}=\frac{\alpha_{s}}{2\pi}\beta+(\frac{\alpha_{s}}{2\pi})^2\frac{\gamma}{9}$,
$U_{II}=\frac{\alpha_{s}}{2\pi}\eta+(\frac{\alpha_{s}}{2\pi})^2\frac{\theta}{9}$,
$V_{I}=(\frac{\alpha_{s}}{2\pi})^2\frac{\zeta}{9}$ and
$V_{II}=(\frac{\alpha_{s}}{2\pi})^2(2n_{f})\frac{\xi}{9}$. The
function $\widetilde{f}(\frac{\delta}{{|{\delta}-1|}})$
($f=G,\Sigma$) is a small constant at $x=0$. At low- $x$, this
constant can be neglected in the Eqs.(15) and (16) due to the
singular behavior of the gluon distribution. On this basis we get:
\begin{eqnarray}
\frac{dG}{dlnQ^{2}}=\tau[U_{I}G({\mu}x)+U_{II}\Sigma({\mu}x)],
\end{eqnarray}
and
\begin{eqnarray}
\frac{d\Sigma}{dlnQ^{2}}=\tau[V_{I}\Sigma({\mu}x)+V_{II}G({\mu}x)],
\end{eqnarray}
where $\tau=\frac{{\delta}^{{\delta}-1}}{{|{\delta}-1|}^{\delta}}$
and $\mu=\frac{\delta}{{|{\delta}-1|}}$. These equations present a
set of formula to extract the gluon distribution function from
singlet structure function and its derivative $d\Sigma/dlnQ^{2}$,
also the singlet structure function from the gluon distribution
and its derivative $dG/dlnQ^{2}$ at small $x$ in the next- to-
leading order of perturbation theory.\\

 Kotikov and Parente [4] presented a set of formula to extracted the
gluon distribution function from the deep inelastic structure
function $ F_{2}$ and its derivative $dF_{2}/dlnQ^{2}$ at small
$x$ in the leading and next to leading order of perturbation
theory. For concrete value of $\delta=0.5$  and the number of
flavors $N_{f}=4$ they have extracted the gluon distribution with
the help this equation:
\begin{eqnarray}
xg(x,Q^{2})=\frac{105}{92e}\frac{1}{\alpha}\frac{1}{(1+26.93\alpha)}[\frac{dF_{2}(x,Q^{2})}{dlnQ^{2}}+\frac{16}{3}\nonumber\\
\alpha(\frac{107}{60}-2ln2)F_{2}(x,Q^{2})+O(\alpha^{2},x^{1-\delta})]
\end{eqnarray}
where $e=\sum_{i}^{f}e_{i}^{2}$ is the sum of squares of quark
charges and $\alpha(Q^{2})=\alpha_{s}(Q^{2})/4\pi$. A different
method for the determination of the gluon distribution at small
values of $x$ has been proposed by Ellis, Kunszt and Levin [24]
based on the solution of the DGLAP evolution equations in the
moment space up to NNLO. In this method the quark and gluon
momentum densities are assumed to behave as $x^{-\omega_{0}}$
where $\omega_{0}$ is a parameter the actual value of which must
be extracted from the data. Here the gluon momentum density for
four flavors is:
\begin{eqnarray}
xg(x,Q^{2})=\frac{18/5}{P^{FG}(\omega_{0})}[\frac{dF_{2}}{dlnQ^{2}}-P^{FF}(\omega_{0})F_{2}],
\end{eqnarray}
where the evolution kernels $P^{FG}$ and $P^{FF}$ calculated in
the $\overline{MS}$ scheme are expanded up to third order in
$\alpha_{s}$.\\

 Applying  Eq.(18), we can arrive at the gluon distribution
 function from the $F_{2}$ proton structure function and its
 scaling violation at low x as the following:
\begin{eqnarray}
xg(x,Q^{2})=\frac{18}{5V_{II}}[\frac{1}{2}\frac{dF_{2}}{dlnQ^{2}}-V_{I}F_{2}].
\end{eqnarray}

By means of these equations we have extracted the gluon
distribution from HERA data, using the slopes $dF_{2}/dlnQ^{2}$
determined in Ref.[21]. Figure 1 shows the extracted values of the
gluon distribution compared to KP model [4], EKL model [24] and
MRST [17,22] parameterization. This result indicate that our
calculations, based upon the available structure functions and its
derivative [21], are
of the same form as the one predicted by the QCD theory.\\

In Regge theory the high energy behavior of hadron-hadron and
photon-hadron total cross section is determined by the pomeron
intercept $\alpha_{P}=1+\delta$, and is given by
$\sigma_{\gamma(h)p}^{tot}(\nu){\sim}\nu^{\delta}$. This behavior
is also valid for a virtual photon for $x<<1$, leading to the well
known behavior,$F_{2}{\sim}x^{-\delta}$, of the structures at
fixed $Q^{2}$ and $x{\rightarrow}0$. The power $\delta$ is found
to be either $\delta=0$ or $\delta=0.5$. The first value
corresponds to the soft Pomeron and the second value the hard
(Lipatov) Pomeron intercept.   The Form $x^{-\delta_{g}}$ for the
gluon parametrization at small $x$ is suggested by Regge behavior,
but whereas the conventional Regge exchange is that of the soft
Pomeron, with $\delta_{g}{\sim}0.0$, one may also allow for a hard
Pomeron with $\delta_{g}{\sim}0.5$. The form $x^{-\delta_{S}}$ in
the sea quark parametrization comes from similar considerations
since, at small $x$, the process
$g{\rightarrow}\hspace{0.1cm}q\overline{q}$ dominates the
evolution of the sea quarks. Hence the fits to early HERA data
have as a constraint $\delta_{S}=\delta_{g}=\delta$, as the value
of $\delta$ should be close to $0.5$ in quite a broad range of low
$x$ [4,9-10,25]. Fig.2 illustrate behavior of the $\tau$ function
in the kinematical region. Derivative of the $\tau$ function is
zero at $\delta=0.5$. For concrete value of $\delta=0.5$ we
obtain:
\begin{equation}
\frac{dG}{dlnQ^{2}}=2[U_{I}G(x)+U_{II}\Sigma(x)],
\end{equation}
and
\begin{equation}
\frac{d\Sigma}{dlnQ^{2}}=2[V_{I}\Sigma(x)+V_{II}G(x)].
\end{equation}

Now let us discuss how the presented results give the independent
evolution equations for the gluon  and singlet structure functions
at low $x$, respectively. By solving these equations, we found:
\begin{widetext}
\begin{eqnarray}
G(x,Q^{2})&=&\frac{1}{2VII}[\frac{1}{2}\frac{d}{dlnQ^{2}}(\frac{1}{UII})\frac{dG(x,Q^{2})}{dlnQ^{2}}+\frac{1}{2UII}\frac{d^{2}G(x,Q^{2})}{dln^{2}Q^{2}}-\frac{d}{dlnQ^{2}}(\frac{UI}{UII})G(x,Q^{2})\nonumber\\
&&-\frac{UI}{UII}\frac{dG(x,Q^{2})}{dlnQ^{2}}]-\frac{VI}{VII}[\frac{1}{2UII}\frac{dG(x,Q^{2})}{dlnQ^{2}}-\frac{UI}{UII}G(x,Q^{2})],
\end{eqnarray}
and
\begin{eqnarray}
\Sigma(x,Q^{2})&=&\frac{1}{2UII}[\frac{1}{2}\frac{d}{dlnQ^{2}}(\frac{1}{VII})\frac{d\Sigma(x,Q^{2})}{dlnQ^{2}}+\frac{1}{2VII}\frac{d^{2}\Sigma(x,Q^{2})}{dln^{2}Q^{2}}-\frac{d}{dlnQ^{2}}(\frac{VI}{VII})\Sigma(x,Q^{2})\nonumber\\
&&-\frac{VI}{VII}\frac{d\Sigma(x,Q^{2})}{dlnQ^{2}}]-\frac{UI}{UII}[\frac{1}{2VII}\frac{d\Sigma(x,Q^{2})}{dlnQ^{2}}-\frac{VI}{VII}\Sigma(x,Q^{2})].
\end{eqnarray}
\end{widetext}

Inserting the effective power behavior corresponding to equation
(11) in these equations gives:
\begin{widetext}
\begin{eqnarray}
\frac{1}{2VII}\frac{1}{2UII}\frac{d^{2}\widetilde{G}(Q^{2})}{dln^{2}Q^{2}}+[\frac{1}{2VII}\frac{1}{2}\frac{d}{dlnQ^{2}}(\frac{1}{UII})-\frac{1}{2VII}\frac{UI}{UII}-\frac{VI}{VII}\frac{1}{2UII}]\frac{d\widetilde{G}(Q^{2})}{dlnQ^{2}}\nonumber\\
+[\frac{VI}{VII}\frac{UI}{UII}-\frac{1}{2VII}\frac{d}{dlnQ^{2}}(\frac{UI}{UII})-1]\widetilde{G}(Q^{2})=0,
\end{eqnarray}
and
\begin{eqnarray}
\frac{1}{2UII}\frac{1}{2VII}\frac{d^{2}\widetilde{\Sigma}(Q^{2})}{dln^{2}Q^{2}}+[\frac{1}{2UII}\frac{1}{2}\frac{d}{dlnQ^{2}}(\frac{1}{VII})-\frac{1}{2UII}\frac{VI}{VII}-\frac{UI}{UII}\frac{1}{2VII}]\frac{d\widetilde{\Sigma}(Q^{2})}{dlnQ^{2}}\nonumber\\
+[\frac{VI}{VII}\frac{UI}{UII}-\frac{1}{2UII}\frac{d}{dlnQ^{2}}(\frac{VI}{VII})-1]\Sigma(Q^{2})=0.
\end{eqnarray}
\end{widetext}
These equations show the structure functions
$\widetilde{f}(Q^{2})$ are  functions of $Q^{2}$. The $lnQ^{2}$
dependence of $\widetilde{f}(Q^{2})$ is observed to be non-linear
[21]. It can be well described by a quadratic expression:
\begin{equation}
\widetilde{f}_{i}(Q^{2})=a_{i}+b_{i}lnQ^{2}+c_{i}(lnQ^{2})^{2},
\hspace{0.5cm} i=g,\Sigma
\end{equation}
where, the function $\widetilde{f}(Q^{2})$ is determined in the
evolution equation resulting from equations (26) and (27) with the
starting parameterizations of partons $Q^{2}=Q_{0}^{2}$ given by
 the input distributions
[10,16-17] of gluon, singlet and its derivatives, respectively.
Therefore, the effective power behavior of the gluon distribution
and the singlet structure function corresponds to:
\begin{equation}
G(x,Q^{2})=(a_{g}+b_{g}lnQ^{2}+c_{g}(lnQ^{2})^{2})x^{(-0.5)},
\end{equation}
and
\begin{equation}
\Sigma(x,Q^{2})=(a_{\Sigma}+b_{\Sigma}lnQ^{2}+c_{\Sigma}(lnQ^{2})^{2})x^{(-0.5)}.
\end{equation}\\

\subsection{3  Results and Discussion}

In this paper, we obtained a new independent evolution
descriptions for the gluon distribution and singlet structure
function based on Regge like behavior of distribution functions
through the equations (23) and (24) respectively.  In these
equations, we need the input functions $F_{2}(x,Q_{0}^{2})$ and
$G(x,Q_{0}^{2})$ and the derivatives of $F_{2}(x,Q_{0}^{2})$ and
$G(x,Q_{0}^{2})$ with respect to $lnQ^{2}$ at each constant $x$
value from the QCD parton distributions in the literature
[10,16-17] . We compared our results of the gluon distribution and
singlet structure function in NLO with MRST2001
 [17], NLO- GRV [16] parameterizations and DL fit [10], respectively.
  We have taken the parameterizations fit to the
 $H1$ data [21] with $x{<}0.1$ and $2{\leq}Q^{2}{\leq}150
 GeV^{2}$. Here we used the QCD cut- off parameter
  $\Lambda^{4}_{\overline{MS}}=0.323\hspace{0.1cm}GeV$
 [17] for
$\alpha_{s}(M_{z^{2}})=0.119$.\\

 In Figs.3-5, we show the prediction of Eqs.26 and 29 for the gluon distribution function. In these calculations
 we need $G(x,Q_{0}^{2})$ and its derivative
with respect to $lnQ^{2}$ at $ Q^{2}=Q_{0}^{2}$. In Fig.3 we
compared our results of  the gluon distribution function with DL
fit [10], $MRSD'_{-}$ [23] and MRST2001 [17] fit. We have taken
the DL parametric form for the starting distribution  at
$Q_{0}^{2}=5 GeV^{2}$  given by
$xg(x,Q^{2})=0.95(Q^{2})^{1+\epsilon_{0}}(1+Q^{2}/0.5)^{-1-\epsilon_{0}/2}x^{-\epsilon_{0}}$
where $\epsilon_{0}$ is equal to $0.437$ according to hard Pomeron
exchange. As it can be seen, the values of the gluon distribution
increase as $x$ decreases but its rate of increment is much higher
than the $MRSD'_{-}$ and MRST  fit. We do however observe that
there is some violation at low $x$.
 This is due to the fact that the hard pomeron exchange defined by DL model is expected to
 hold in the low $x$ limit. One can see that in this case the scaling
 with DL fit is nearly preserved.\\

To illustrate better our calculations at low $x$ we have plotted
$G(x)$ verses $x$ variable [see Fig.4]. One clearly sees that our
results increases when $x$ decreases, but with a somewhat smaller
rate. In this figure, we take the NLO-GRV fit [16] input gluon
density at $Q_{0}^{2}=1 GeV^{2}$ and compared our results with GRV
fit, $MRSD'_{-}$ [23] and MRST2001 [17] fit. For $Q^{2}$ constant,
there is a cross-over point for both of the curves whose
predictions are numerically equal. The cross-over point shifts to
MRSD$^{'}_{-}$ [23] as $x$ decreases. However, we see that this
behavior is due to the fact that the our calculations are
dependent to the input conditions.\\

In Fig.5 we present the gluon distribution $G(x)$ for the H1 HERA
proton parameterization at $Q^{2}=20 GeV^{2}$ [21] for different
low-$x$ values. The initial condition for the evolution of the
gluon density is assumed to be of the form $xg(x,Q_{0}^{2})=
1.1x^{(-0.247)}(1-x)^{17.5}(1-4.83\sqrt{x}+68.2x)$  for
$Q^{2}{\geq}3.5 GeV^{2}$ at the initial scale $Q_{0}^{2}=4
GeV^{2}$. The gluon distribution $G(x)$ is increasing when $x$ is
decreasing. In the same graph we present the $G(x)$ values for the
H1 [21] data, $MRSD'_{-}$ [23] and MRST2001 [17] global fit
results; but its rate of increment is higher than MRST and smaller
than $MRSD'_{-}$. Our results show that the calculations are
sensitive  to the initial conditions at $Q^{2}=Q_{0}^{2}$. For any
initial condition the figures show good agreement between our
results and  those parameterizations at low $x$. We show, in this
figure, the best fit with the MRST gluon distribution
parameterization
corresponding to the initial condition H$_{1}$ data.\\

In Fig.6, we show the prediction of Eqs.27 and 30 for the singlet
structure function. We obtain our results with the input
parameterization  at the initial scale $Q_{0}^{2}=5 GeV^{2}$ and
 compared with the DL fit [10], MRST2001 [17] fit and H$_{1}$
data [21] with the total errors at $Q^{2}=20 GeV^{2}$.  In this
figure we observe a continuous rise towards low $x$. The $lnQ^{2}$
dependence of $F_{2}$ is observed to be non- linear. It can be
well described by a quadratic expression:
\begin{equation}
\widetilde{\Sigma}(Q^{2})=a_{S}+b_{S}lnQ^{2}+c_{S}(lnQ^{2})^{2},
\end{equation}
which nearly coincides with the QCD fits in the kinematic range of
this calculation. Then the effective power behavior of the singlet
structure function corresponds to:
\begin{equation}
F_{2}(x,Q^{2})=\widetilde{F_{2}}(Q^{2})x^{(-0.5)}.
\end{equation}
 This behavior is  associated with the exchange of an object known
 as the hard Pomeron. Donnachie and Landshoff [9-10] shows this behavior by the simplest fit
 to the small-$x$ data corresponds to:
 \begin{equation}
F_{2}(x,Q^{2})=\sum_{i=0,1}f_{i}(Q^{2})x^{-\epsilon_{i}},
\end{equation}
where the $i=0$ term is hard pomeron exchange and $i=1$ is soft
pomeron exchange. These parameters obtained from the best fit to
all the small- $x$ data for $F_{2}(x,Q^{2})$ together with the
data for $\sigma^{{\gamma}p}$. So that our structure function is
dominant
 at small $x$ by hard Pomeron exchange. This powerful approach to the
 small-$x$ data for $F_{2}(x,Q^{2})$ is to extend the Regge
 phenomenology that is so successful for hadronic processes [7].
 Regge theory relates high-energy behavior to singularities in the
 complex angular momentum plane [6]. So, for deep inelastic
 scattering, the soft Pomeron contributions is not sufficient to
 describe the rapid rise with $1/x$ seen in the data at small $x$
 and large $Q^{2}$. This singularity is hard Pomeron [9,10].\\

In conclusion, a set of new formulae connecting the gluon density
with its derivative and the singlet structure function with its
derivative with respect to $lnQ^{2}$ at low $x$ have been
presented. We found that one can use Regge theory to constrain the
initial parton densities at $Q^{2}=Q_{0}^{2}$ and obtain the
distributions at higher virtualities with the DGLAP evolution
equation. Careful investigation of our results shows a good
agreement with the previous published parton distributions based
upon QCD. The gluon distribution and singlet structure functions
will increase as usual, when $x$ decreases. The form of obtained
distribution functions
 for the gluon distribution and the singlet structure functions  are similar to the one predicted
 from parton parameterization. The formulae were used to generate
 the parton distributions are in agreement with the rise observed by
 H1 experiments. We observed a continuous rise towards low $x$. The $lnQ^{2}$
dependence of $f(x,Q^{2})$ is observed to be non- linear by a
quadratic expression:
\begin{equation}
\widetilde{f}(Q^{2})=a_{i}+b_{i}lnQ^{2}+c_{i}(lnQ^{2})^{2},
(i={g}\hspace{0.2cm} or\hspace{0.2cm} \Sigma)
\end{equation}
which nearly coincides with the QCD fits in the kinematics range
of these calculations. Thus the effective power behavior of the
 parton densities corresponds to:
\begin{equation}
f(x,Q^{2})=\widetilde{f}(Q^{2})x^{(-0.5)}.
\end{equation}
This behavior is associated with the exchange of an object known
 as the hard Pomeron at small $x$. The obtained results give strong indications that the proposed
formulae, being very simple, provides relatively accurate values
for the gluon distribution and structure function.\\
\newpage

\textbf{Figure Captions}\\
Fig.1. The solid circles represent our gluon prediction (Eq.21)
using the structure function $F_{2}$ and $dF_{2}/dlnQ^{2}$ are
taken by the H1 [21] collaboration for a range of $x$ values at
$Q^{2}=20 GeV^{2}$. The error bar show total errors to H1 data. We
compared our results with KP model [4], EKL model [24] and MRST
fit[17,22](Solid line).\\
Fig.2. Behavior of the $\tau$ function into $\delta$ values.\\
Fig.3. The gluon distribution given by Eqs.(26) and (29) against
$x$ at fixed $Q^{2}=20 GeV^{2}$ value and compared with DL
fit[10](Solid line), MRSD$'_{-}$[23](Dot line) and MRST
fit[17](Dash line). The starting parameterization of the gluon
density at $Q_{0}^{2}=5 GeV^{2}$  given by the DL model.\\
Fig.4. The gluon distribution given by Eqs.(26) and (29) against
$x$ at  fixed $Q^{2}=20 GeV^{2}$ value and compared with
NLO-GRV[16](Solid line), MRSD$'_{-}$[23](Dash line) and MRST
fit[17](Dot line). The starting parameterization of the gluon
density at $Q_{0}^{2}=1 GeV^{2}$  given by the NLO-GRV.\\
Fig.5. The gluon distribution given by Eqs.(26) and (29) against
$x$ at  fixed $Q^{2}=20 GeV^{2}$ value and compared with H1[21]
data, MRSD$'_{-}$[23](Dash line) and MRST fit[17](Solid line). The
starting parameterization of the gluon density at $Q_{0}^{2}=4
GeV^{2}$  given by the H1 Collaboration data.\\
Fig.6. The calculated values of the singlet structure function
$F_{2}(x,Q^{2})$ plotted as functions of $x$ by Eqs.(27) and (30)
into the starting parameterization of the structure function at
$Q_{0}^{2}=5 GeV^{2}$  given by the DL model respectively,
compared with NLO QCD fit to the $H1$ data with total errors  [21]
also with the DL fit [10](Solid line) and the singlet structure
function MRST fit.\\

\subsection{References}
1. Yu. L.Dokshitzer, Sov.Phys.JETPG {\bf6}, 641(1977 );
G.Altarelli and
G.Parisi, Nucl.Phys.B{\bf126}, 298(1997 ); V.N.Gribov and L.N.Lipatov, Sov.J.Nucl.Phys.{\bf28}, 822(1978).\\
2. L.F.abbott, W.B.Atwood and A.M.Barnett, Phys.Rev.D {\bf22}, 582(1980).\\
3. A.M.Cooper- Sarkar, R.C.E.Devenish and A.DeRoeck, Int.J.Mod.Phys.A {\bf13}, 3385( 1998 ).\\
4. A.K.Kotikov  and G.Parente, Phys.Lett.B {\bf379}, 195(1996 ); J.Kwiecinski, hep-ph/9607221.\\
5. R.D.Ball and  S.Forte, Phys.Lett.B {\bf335},  77(1994 )\\; Phys.Lett.B {\bf336}, 77(1994 ).\\
6.  P.D.Collins, An introduction to Regge theory and
high-energy physics(Cambridge University Press,1997).\\
7. A.Donnachie and  P.V.Landshoff, Phys.Lett.B {\bf296}, 257(1992).\\
8. E.A.Kuraev, L.N.Lipatov and V.S.Fadin, Sov.Phys.JETP {\bf{44}}, 443(1976);\\
 Sov.Phys.JETP {\textbf{45}}, 199(1977);\\
 Y.Y.Balitsky and L.N.Lipatov, Sov.Journ.Nucl.Phys. {\textbf{28}}, 822(1978).\\
 9. A.Donnachie and P.V.Landshoff, Phys.Lett.B {\bf437}, 408(1998 ).\\
 10. A.Donnachie and P.V.Landshoff, Phys.Lett.B {\bf550}, 160(2002 )\\; P.V.Landshoff,hep-ph/0203084.\\
11. P.Desgrolard, M.Giffon, E.Martynov and E.Predazzi, Eur.Phys.J.C {\bf18}, 555(2001).\\
12. P.Desgrolard, M.Giffon and E.Martynov, Eur.Phys.J.C {\bf7},  655(1999).\\
13. M.B.Gay Ducati  and V.P.B.Goncalves, Phys.Lett.B {\bf390}, 401(1997).\\
14. K.Pretz, Phys.Lett.B {\bf311}, 286(1993); Phys.Lett.B {\bf332}, 393(1994).\\
15. A.V.Kotikov, hep-ph/9507320.\\
16. M.Gluk, E.Reya  and A.Vogt, Z.Phys.C {\bf67}, 433(1995 ); Euro.Phys.J.C {\bf5}, 461(1998 ).\\
17. A.D.Martin, R.G.Roberts, W.J.Stirling and R.S.Thone, Phys.Lett.B {\bf531}, 216(2002).\\
18. W.Furmanski  and R.Petronzio, Phys.Lett.B {\bf97}, 437(1980 ); Z.Phys.C {\bf11}, 293(1982).\\
19. R.K.Ellis , W.J.Stirling and B.R.Webber, QCD and Collider
Physics(Cambridge University Press,1996).\\
20. R.G.Roberts, The structure of the proton(Cambridge University Press,1990).\\
21. C.Adloff, et.al, $H1$ Collab. Eur.Phys.J.C {\bf21}, 33(2001 ).\\
22. A.D.Martin, R.G.Roberts, W.J.Stirling and R.S.Thone, Eur.Phys.J.C {\bf23}, 73(2002);\\
A.Vogt, S.Moch  and A.M.Vermasern,  Nucl.Phys.B {\bf691}, 129(2004).\\
23. A.D.Martin, R.G.Roberts and W.J.Stirling, Phys.Lett.B {\bf354}, 155(1995); Phys.Lett.B {\bf306}, 145(1993).\\
24. R.K.Ellis, Z.Kunszt and E.M.Levin, Nucl.Phys.B {\bf420}, 517(1994).\\
25. A.D.Martin, M.G.Ryskin and G.Watt, arXiv:hep-ph/0406225(2004);
J.Kwiecinski and A.M.Stasto, Phys.Rev.D {\bf66}, 014013(2002).\\
\newpage
\begin{figure}
\includegraphics[width=1\textwidth]{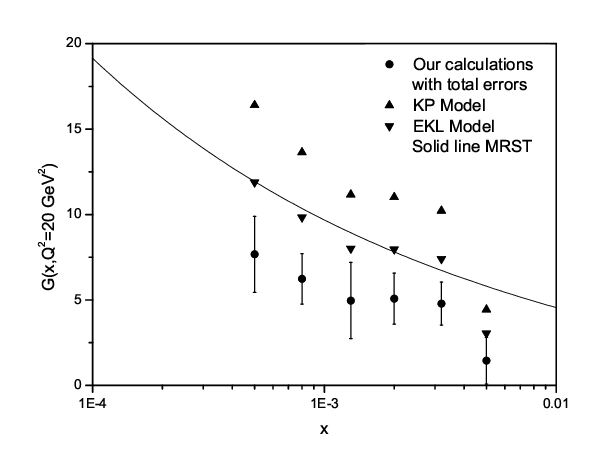}
Fig.1
\end{figure}
\begin{figure}
\includegraphics[width=1\textwidth]{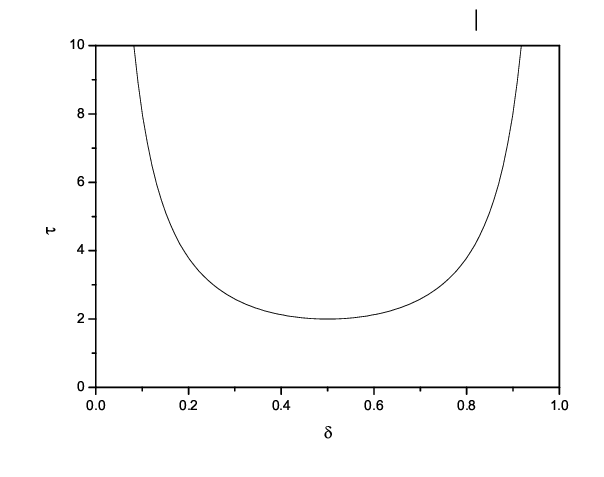}
Fig.2
\end{figure}
\begin{figure}
\includegraphics[width=1\textwidth]{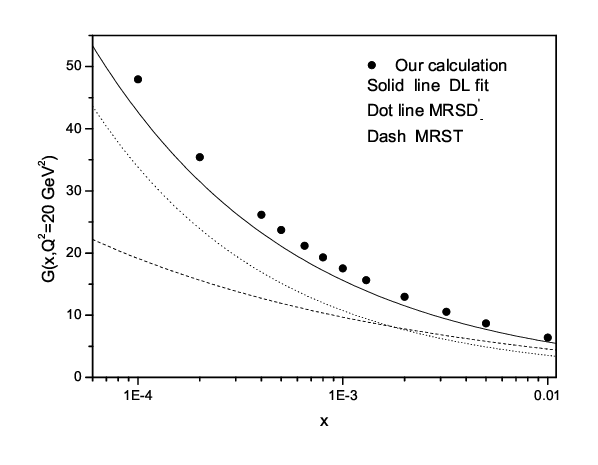}
Fig.3
\end{figure}
\begin{figure}
\includegraphics[width=1\textwidth]{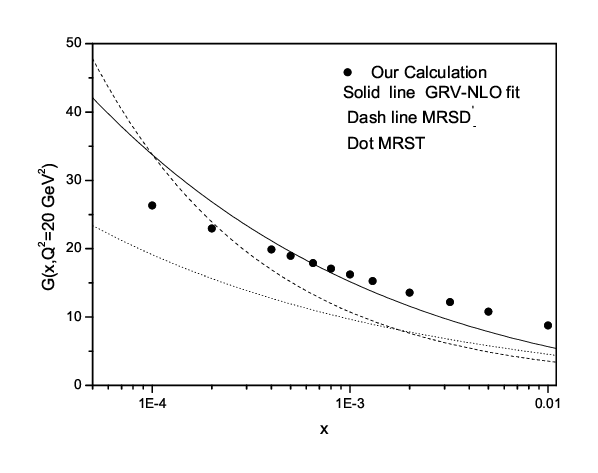}
Fig.4
\end{figure}
\begin{figure}
\includegraphics[width=1\textwidth]{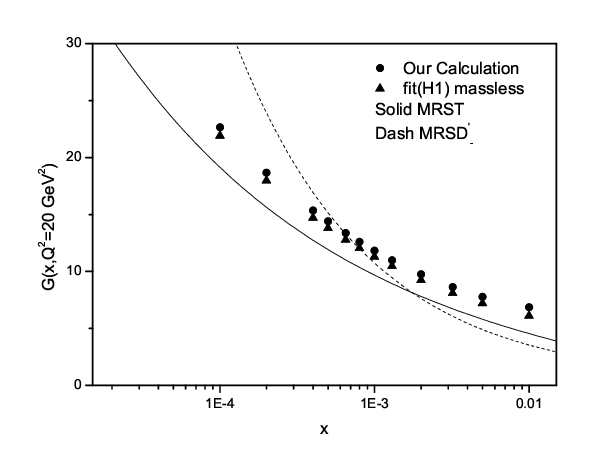}
Fig.5
\end{figure}
\begin{figure}
\includegraphics[width=1\textwidth]{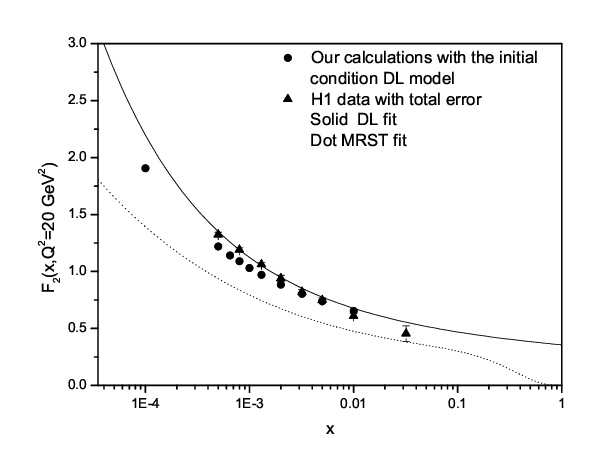}
Fig.6
\end{figure}
\end{document}